\documentclass[prl,twocolumn,aps,amssymb,footinbib]{revtex4}
\usepackage{amssymb}

\usepackage{amssymb}

\usepackage{graphicx}
\usepackage{amsmath}
\usepackage{times}

\begin{document}

\title{Noise Correlations and Quantum Coherence in Hard-core Bosons in One-dimensional Lattices}
\author{ Ana Maria Rey \cite{Ana} , Indubala I Satija \cite{Indu} and Charles W Clark, }
\affiliation{ National Institute of Standard and Technology, Gaithersburg MD, 20899}
\date{\today}

\begin{abstract}
Noise correlations, such as those observable in the time of flight
images of a released cloud, are calculated for hard-core
bosonic (HCB) atoms.
We find that
the standard  mapping of HCB systems onto spin-$1/2$ XY models
fails in application to computation of noise correlations. This is
due to the contribution of {\it multiply occupied virtual states}
to noise correlations in bosonic systems. 
Such states do not exist in spin models.
We use these correlations
to explore quantum coherence of the ground states and re-address the relationship
between the peaks present in noise correlation and the Mott phase.
Our analysis points to distinctive new experimental signatures of the Mott phase.
The importance of
these correlations is illustrated in an example of
a quasiperiodic potential that exhibits a localization transition. In
this case, in contrast to the momentum distribution,
the noise correlations reveal the presence of quasiperiodic order in the
localized phase.

\end{abstract}
\maketitle

In recent years, great experimental progress has been achieved in
the coherent control of ultra-cold gases. In particular by loading a
Bose-Einstein condensate into a tight two-dimensional optical
lattice, an array of one dimensional tubes has been created
\cite{Tolra,Weiss,Paredes,Moritz,Fertig}. Using this setup, recent
experiments have been able to successfully enter the Tonks-Girardeau
regime (TG) \cite{Tolra,Weiss,Paredes}, where the strong
interactions between the bosons mimic the Pauli exclusion principle
\cite{GR}, and to realize the Mott insulator transition in one
dimension\cite{Paredes}.

Recent theoretical and experimental studies have shown that the
atomic shot noise \cite{Altman,Greiner,Foelling,New} in the time of
flight images can be useful to decode various correlations
underlying many-body states of trapped ultra-cold atoms. For
ultra-cold bosons in a Mott insulator state, for example, these
second order correlations have been proven to complement the
standard, momentum-distribution based, characterization of the phase coherence.
\cite{Altman,Foelling}. Such correlations have also
been used to probe  pair condensation in a fermionic superfluid
\cite{Greiner}.

In this paper, we develop a theoretical framework to compute noise
correlations in one dimensional hard core bosonic atoms (HCB) in a lattice.
The HCB Hamiltonian is identical to that of spin-$1/2$ XY model
which in turn can be mapped to that of spinless fermions.\cite{Lieb,LM}
Although these three systems
have identical spectra and local observable, their
off-diagonal correlation functions differ. Experimentally relevant
two-point correlation functions are identical for HCB and spin-$1/2$ systems,
and the general formulation to calculate these
functions  was developed by Lieb and Mattis\cite{LM}, based on Wick's theorem.
To the best of our knowledge, this formulation
has not been extended to higher order correlations. Here we
complete such an extension to treat
four-point correlation functions. One of the central results of this
paper is the discovery of important differences between higher
order correlation functions of HCB and spin-$1/2$ systems. The root
of this difference is the fact that HCB systems have virtual states
which may be multiply occupied, whereas
spin-$1/2$ states are always at most singly occupied.

We  use the noise correlations  to explore quantum coherence in
systems  with and without an external parabolic  confinement. Our
analysis generalizes, to the strongly correlated (fermionized)
regime,
 previous studies of noise correlations
carried out in
the Mott insulator limit.\cite{Altman,Foelling}
We show that,
{\it independent of the filling factor of the system}, HCB
exhibits second order coherence displayed  as peaks in the noise
shot images which reflect the order induced by the lattice potential.
This suggests that the second order coherence
in noise correlations is a generic attribute of the strongly correlated regime and does
not merely indicate reduced number fluctuations (This was also noted in earlier
studies in a different context.\cite{Altman,New}).
On the other hand, we find that the intensity of peaks
depends exclusively on the
relative coordinates only in Mott-insulator limit. Thus a regular pattern in  the noise
correlation could serve as a definitive signature of the existence of a Mott
phase. Finally, we show that noise correlations are also important
in describing the quantum coherence of HCB systems
in the presence of quasiperiodic disorder.
Quasiperiodicity induces an Anderson-type localization transition
\cite{QP} in the system.
In the delocalized phase, both the first and the second order
correlations are found to exhibit characteristic Bragg peaks at the
Fibonacci sites reflecting quasiperiodic order. However, in the localized phase only noise
correlations show these peaks.

The Bose-Hubbard Hamiltonian describes bosons in optical
lattices  when the lattice is loaded in such a way that only the
lowest vibrational level of each lattice site is occupied and
tunneling occurs only between nearest-neighbor sites,\cite{Jaksch}:
\begin{equation}
H= -J\sum_{\langle i,j\rangle
}\hat{a}_i^{\dagger}\hat{a}_{j}+\frac{U}{2}\sum_{j}\hat{n}_j(\hat{n}_j-1)
+ \sum_{j} V_j \hat{n}_j
\end{equation}
\noindent Here $\hat{a}_j$ is the bosonic
annihilation operator of a particle at site $j$,
$\hat{n}_j=\hat{a}_j^{\dagger}\hat{a}_{j}$, and the sum $\langle
i,j\rangle$ is over nearest neighbors. The hopping parameter $J$,
and the on-site interaction energy  $U$ are functions of the lattice
depth. $V_j$ represents any other external potential such as a
parabolic confinement or on-site disorder.

In a typical experiment,
atoms are released by turning off the external potentials at time $t=0$.
The atomic cloud expands
, and is photographed after it enters the ballistic regime.
Assuming that the atoms are noninteracting from the time of release,
properties of the initial state can be inferred from the spatial images:
$\langle \hat{n}[x(t)]\rangle$, reflects the initial
momentum distribution, $n_q$,
and the image shot noise, $\mathcal{G}[x(t),x'(t)]$
reflects the momentum space fluctuations,
namely the {\it noise correlations}, $\Delta(q_1,q_2)$,

\begin{eqnarray*}
\langle \hat{n}[x(t)]\rangle &\propto&  \langle \hat{n}_q\rangle
=\sum_{n,m} e ^{i\frac{2\pi}{L} q (n-m) } \langle \hat{a}_n^\dagger
\hat{a}_m\rangle\\
\mathcal{G}[x(t),x'(t)] &\propto&  \langle (\hat{n}_{q_1}-\langle
\hat{n}_{q_1}\rangle)( \hat{n}_{q_1}- \langle
\hat{n}_{q_1}\rangle)\rangle \equiv \Delta(q_1,q_2)
\\&=&\sum_{n,m,l,j} e ^{i \frac{2\pi}{L}{q}_1 (n-m) } e ^{i \frac{2\pi}{L}{q}_2 (i-j) } \langle
\hat{a}_n^\dagger \hat{a}_m \hat{a}_l^\dagger \hat{a}_j\rangle
\notag
\end{eqnarray*}

\noindent where $L$ is the number of lattice sites.
In the strongly correlated  regime, Eq. (1) can be replaced by
the HCB Hamiltonian,
\begin{equation}
H-J\sum_j(\hat{b}_j^{\dagger}\hat{b}_{j+1}
+\hat{b}_{j+1}^{\dagger}\hat{b}_{j})+\sum_{j} V_j \hat{n}_j
\end{equation}
Here $\hat{b}_j$ is the annihilation operator at the lattice  site $j$
which satisfies
$[\hat{b}_{i\neq j},\hat{b}_j^{\dagger}]=0$, and the on-site condition
${\hat{b}_j}^2={\hat{b}_j^{\dagger}}{}^2=0$, which
suppresses multiple occupancy of lattice
sites. The same relations are fulfilled by
Spin-$1/2$ raising and lowering operators.
However, the exact on-site commutation relation differ between HCB and
spin-$1/2$ operators, and this becomes important in processes involving
virtual states.
This distinction between
spin and HCB models has not been explored in  earlier studies;
it has no effect on either the
mapping of HCB to
free fermions or on the calculations of local observable and the
momentum distribution. However, the presence of multiply occupied lattice sites
in virtual states
that occurs in HCB strongly affects the noise correlation.
A simple example of this can be seen
in the computation of the correlation function,
$\langle 1| \hat{b}\hat{b}^{\dagger}|1\rangle$.
In the spin-$1/2$ model  this correlation function is zero, while for  HCB it
is equal to $2$.
On the other hand
$\langle 1| \hat{b}^{\dagger}\hat{b}|1\rangle=1$ for both systems.

We now describe the way in which we have generalized the approach of
Lieb and Mattis. We have found a simple recipe to take into account
the problem of multiple occupancy of the virtual state. In our
calculation of the four point correlations $\langle
\hat{b}_n^\dagger \hat{b}_m \hat{b}_l^\dagger \hat{b}_j\rangle$,
each occurrence of  a term
 $\hat{b}_j
\hat{b}_j^\dagger$ is replaced  by the  $1+ \hat{b}_j^\dagger
\hat{b}_j$ (Note that no new rules are needed if we encounter number
operators, $\hat{b}_j^\dagger \hat{b}_j$ or operators at different
sites). In other words, only those four-point correlation functions
involving two or more equal sites where the pair has the form
$\hat{b}_j \hat{b}_j^\dagger$, should be treated differently from
that of the corresponding spin correlations. We will refer this
recipe as {\it multiple occupancy of virtual state rule} (MOV). The
validity of MOV was checked by comparing various correlation
functions obtained using the above recipe with those obtained by
diagonalizing a full Bose-Hubbard Hamiltonian with a large $U$
value.

The following procedure is used  to calculate the four-point
correlation functions. We first rearrange the operators so that the
site index is ordered (this is only relevant for the case when three
or more site indices are different); then we apply MOV; next, we use
the Jordan-Wigner transformation accordingly to the prescription of
Lieb and Mattis; and finally we use Wick's theorem to write higher
order correlations in terms of the free-fermionic propagators,
$g_{lm}=\sum_{s=0}^{N-1} \psi_l^{*(s)} \psi_m^{(s)}$, with $N$ being
the total number of atoms and $ \psi_l^{(s)}$  the $s^{th}$
eigenfunctions of the single-particle Hamiltonian
$-J(\psi_{l+1}^{(s)}+\psi_{l-1}^{(s)})+V_l \psi_{l}^{(s)}=E^{(s)}
\psi_{l}^{(s)}$. To present our results we denote the creation and
the annihilation operators by $\hat{b^{\alpha}}$ , where $\alpha
=+1(-1)$ for annihilation (creation) operators, respectively, and
the "site ordered" four-point correlation function is designated by
$\chi^{\alpha\beta\gamma\delta}_{abcd}$.

\begin{equation}
 \langle
\hat{b}_n^\dagger \hat{b}_m \hat{b}_l^\dagger \hat{b}_j\rangle
\mapsto \langle \hat{b}_{a}^{(\alpha)} \hat{b}_{b}^{(\beta)}
\hat{b}_{c}^{(\gamma)} \hat{b}_{d}^{(\delta)}\rangle \equiv
\chi^{\alpha\beta\gamma\delta}_{abcd}
\end{equation}

In this equation it is implicit that $a\leq b \leq c \leq d$, and
this order is implied in all the expressions that follow. We define
$G_{ij}\equiv 2 g_{ij} -\delta_{i,j}$ and $B_{ij}\equiv \langle
\hat{b}_{i}^{(\dagger)} \hat{b}_{j}\rangle$. The latter can be
calculated in terms of $G_{ij}$ \cite{LM}. We introduce four
matrices ${\bf M},{\bf S},{\bf X}$ and ${\bf Y}$  in terms of which
our results for the correlation functions can be written. These
matrices are presented at the end of this paper due to their
notational complexity. Our results for the correlation functions
then take the form:

\begin{eqnarray}
\chi^{\alpha\beta\gamma\delta}_{abbd}&=&\frac{1- \beta
\gamma}{2}\left[\frac{1}{4} |{\bf M}(a,b,d
)|+\left(\frac{1}{2}+\delta_{\beta,-1}\right) B_{ad}
\right],\notag\\
 \chi^{\alpha\beta\gamma\delta}_{aacd}&=&\frac{1- \alpha\beta}{2}\left[\frac{1}{4} |{\bf
S}(a,c,d)|+\left(\frac{1}{2}+\delta_{\alpha,-1}\right)
B_{cd}\right],\notag\\
\chi^{\alpha\beta\gamma\delta}_{abcc}&=&\frac{1- \gamma
\delta}{2}\left[\frac{1}{4} |{\bf
S}(c,a,b)|+\left(\frac{1}{2}+\delta_{\gamma,-1}\right)
B_{ab}\right],\notag\\
\chi^{\alpha \beta \gamma \delta}_{abcd}&=&(-1)^
{b+d-c-a}\left[\frac{2- \gamma\delta- \alpha\beta }{16} |{\bf
X}(a,b,c,d )|+ \right]\notag\\
&& \left. \frac{\beta}{4}(\delta_{\gamma,-1}-\delta_{\alpha,-1})
|{\bf Y}(a,b,c,d )|\right].
\end{eqnarray}

When three, four or two pairs of indices are equal the calculation
is rather straight forward. The non-vanishing correlation functions
of this type are given by: $ \langle \hat{b}_n^\dagger\hat{b}_n
\hat{b}_n^\dagger \hat{b}_n\rangle=g_{nn}$,
$\langle\hat{b}_n^\dagger \hat{b}_n\hat{b}_m^\dagger
\hat{b}_m\rangle=g_{nn} g_{mm}- g_{nm}^2$, $ \langle
\hat{b}_n^\dagger \hat{b}_m\hat{b}_m^\dagger \hat{b}_n\rangle=
g_{nn} g_{mm}- g_{nm}^2+ g_{nn}$ and $\langle\hat{b}_n^\dagger
\hat{b}_n \hat{b}_n^\dagger \hat{b}_m\rangle=\langle
\hat{b}_m^\dagger \hat{b}_n \hat{b}_n^\dagger
\hat{b}_n\rangle=B_{nm} $ with
$n\neq m$.

The  formulas  above were used in numerical calculations of the
momentum distributions and the noise correlations for lattices of
$L=55$ sites and $N=19$ atoms with periodic boundary conditions.
Fig.1 shows both first and second order correlations along with the
density profile . There we also include  effects of a magnetic
confinement $V_j=j^2 \Omega/J $ on various correlations.
Experimentally the ratio $\Omega/J$ can be changed either by
changing the lattice depth or the external magnetic confinement. The
parameters used in our analysis relate to typical experimental
set-ups such  as the ones reported in ref\cite{Paredes}.

The density profiles (Fig1a) for different values of $\Omega/J$ show that
only in the case $\Omega/J=0.17$ does the ground state of the system
correspond to a Mott insulator. In this case all the central $N$
sites have unit filling. For $\Omega/J=0.018 $ localization takes
place only at the central site, while for $\Omega/J=0.008$ and $0$,
all the sites have filling factor less than unity. The formation of
a Mott state with reduced number fluctuations for $\Omega/J=0.17$ is
clearly signaled by the momentum distribution (Fig1b), which shows a flat
profile. On the other hand for the other cases where number
fluctuations are important, there is a clear peak in the momentum
distribution at $q=0$. This peak reflects the quasi-long-range
correlations  in the density matrix of these systems \cite{Rigol}.

In contrast to the momentum distribution, the noise-correlations 
$\Delta(q_1,0)$ (Fig1c) for all values of  $\Omega/J$ are peaked at $q_1=0$.
This  lattice induced peak is  observable due to the strong particle
correlations in
the fermionized regime
and disappears in the weakly correlated regime
where most of the atoms are Bose-condensed. In our numerical study
without any confining potential, where we vary the filling factors
(not shown in the figures), the central ($q_1=0$) peak in the noise
correlations showed the same particle-hole symmetry as the central
peak in the momentum distribution. The value of $\Delta(0,0)$
increases with filling  factor, $\nu$, up to $\nu=1/2$, and then
decreases to  its minimum value of $2-1/L$ at unit filling when the
system is in the Mott phase. For partially filled states ($N < L$),
the quantity $\Delta(q_1,0)$ has a dip near $q_1=0$. This satellite
dip disappears in the Mott phase,  in which case, $\Delta(q_1\neq
0,0)$ is a constant equal to $-1/L$. As seen in Fig.~1, these
observations  remain qualitatively valid in the presence of a trap,
where the analysis of the noise correlations is more complicated,

\begin{figure}[htbp]
\includegraphics[width=3. in,height=3.3 in]{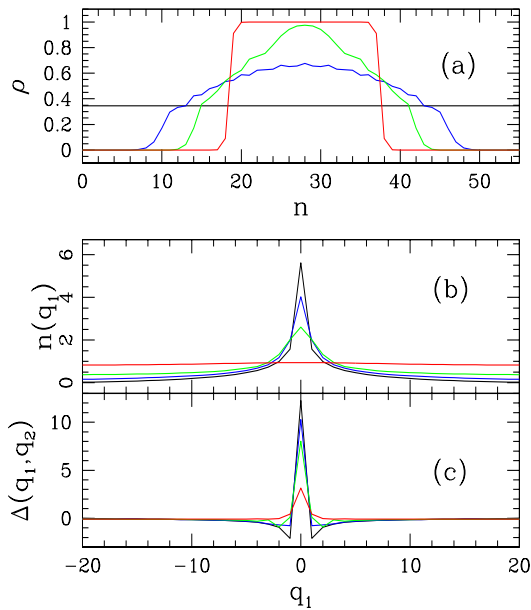}
\leavevmode
\caption{ Density (a), momentum distribution (b) and noise correlation (c) with different trapping potentials: $\Omega/J=$
$0$(black), $0.008$(blue), $0.018$(green), $0.17$(red)). Here $q_2=0$, $N=19$ and $L=55$. For finite $\Omega$,
the correlation functions are renormalized by a scaling factor $N/Z$ where $Z$ are the number of sites with
non-zero density.}
\label{fig1}
\end{figure}

\begin{figure}[htbp]
\includegraphics[width=3.2in,height=2.4in]{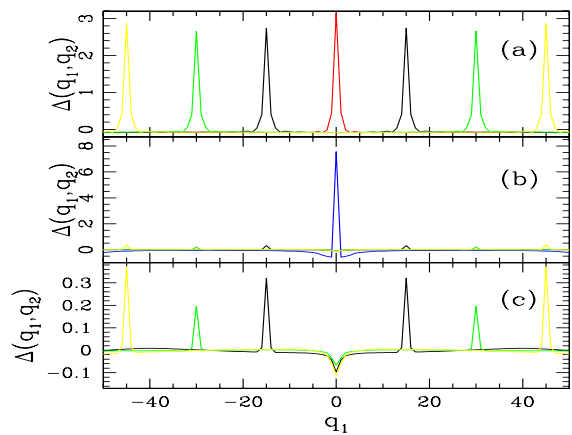}
\leavevmode
\caption{ Mott (a) and non-Mott(b) phase (obtained with $\Omega/J=0.17$ and $\Omega/J=0.008$ respectively)
using $L=55$ and $N=19$. Each color corresponds to $\Delta(q_1,q_2)$ for a fixed $q_2$ as a function of $q_1$.
Thus different colors correspond to different values of $q_2$.
Peaks occur when $q_1=q_2$.
Red (blue) show the peak when $q_2=0$ in the Mott ( non-Mott) phase.
Other peaks correspond to
$q_2=15$(black), $30$(green), $45$(yellow). Panel (c) is a blowup of panel (b) where
we show $q_2 \ne 0$ correlations only.}
\label{fig2}
\end{figure}

The existence of second order coherence in HCB systems, independent
of their  filling factor, implies that the peaks cannot be used as a
signature of the Mott insulator in 1D systems. Nevertheless, our
numerical calculations show that only when the system is a Mott
insulator, the noise-correlations exhibit a regular pattern, {\it
i.e.} $\Delta(q_1,q_2)\approx \Delta(q_1-q_2)$ (small differences
seen in Fig.~2 are due to the finite trap). Fig.~2 illustrates the
contrast between the Mott (Fig2a) and the non-Mott (Fig2b-c) phase.
In addition to the variation in the intensity of the peaks, the
$q_2 \ne 0$ correlations show a dip at the center ( Fig2c).
\begin{figure}[htbp]
\includegraphics[width=3.2in,height=2.7in]{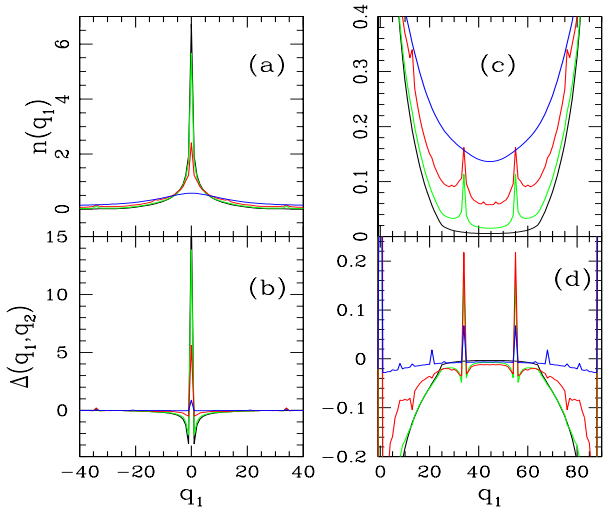}
\leavevmode
\caption{Quantum coherence as quasiperiodic disorder varies:
$\lambda=0$(black), $0.5$(green), 
$1$(red), $2$ (blue).
Figures on the right show a blowup of the Fibonacci peaks which are barely visible on the left.
Additionally, distinct harmonic peaks can be seen
at $68,76,81$. (Here $q_2=\Omega/J=0$)}
\label{fig3}
\end{figure}

As a final example  we introduce quasiperiodicity by adding a
potential $V_j=2\lambda \cos(2\pi\gamma j+\phi)$ where $\lambda$ is
an amplitude and $\gamma$ is  an irrational number which introduces
competing periodicities in the system. We choose
$\gamma=(\sqrt(5)-1)/2$ and $\phi=\pi/4$. The free-fermion
Hamiltonian with this potential exhibits localization-delocalization
transition at $\lambda=1$ \cite{QP}. In our numerical study,
$\gamma$ is replaced by a ratio of two Fibonacci numbers
$F_n/F_{n+1}$, ($F_1=F_0=1, F_{n+1}=F_{n}+F_{n-1}$), which describe
the best rational approximant, obtained by continued fraction
expansion of $\gamma$ \cite{QP}. Here we treat the case of
$\gamma=55/89$, $L=89$ and $N=25$. In the extended phase
($\lambda<1$), the spectrum is effectively continuous and the
eigenstates are of the Bloch type, while in the localized phase, the
spectrum is point-like and the eigenstates are exponentially
localized.  Because in its ground state, all the lowest $N$ single
particle levels are occupied,  the HCB system displays its
metal-insulator transition  in the density profile: in the extended
phase, the on-site density varies smoothly between sites, while in
the localized phase it is discontinuous.

The effects of quasiperiodic disorder on HCB are depicted in Fig.3.
We see in Fig.3a  that the localization transition destroys the cusp
in the central peak of the momentum distribution, demonstrating the
loss of first order quantum coherence. However, second order quantum
coherence is preserved, as shown in  the structure of the noise
correlations, Fig.3b. This is analogous to the changes in coherence
associated with  the Mott transition in the absence of disorder.
Additional peaks reflecting quasiperiodic order of the system appear
at the reciprocal lattice vectors $\frac{2\pi}{L}F_n$ (Figs.3c,d).
The localization transition destroys these peaks in the momentum
distribution (3c), yet they remain in the noise correlations (3d).
In other words, in the localized phase, no trace of the
quasiperiodic order remains in first order correlations, but  it is
still seen clearly in second order correlations. The intensities  of
these second order Bragg peaks increase with disorder in the
extended phase and reach a maximum value at the onset to
localization, after which they decrease. Furthermore, at the
critical point as well as in the localized phase, we see  additional
structure at various harmonics of two frequencies that underly the
quasiperiodicity of the  system. The harmonic peaks in the localized
phase reflect self-similar fluctuations of the wave function
\cite{KS}. We hope that our studies of quasiperiodic quantum
coherence (whose details will be published elsewhere) will stimulate
experimental studies of such
 systems, for example in two-color optical  superlattices \cite{Roth}.

\begin{widetext}
\begin{eqnarray*}
{\bf M}=\left(\begin{array}{ccccc}
  G_{aa+1} & G_{ab-1} & G_{ab+1} & .. & G_{ac} \\
  \vdots &  & & & \vdots \\
  G_{b-1a+1} & G_{b-1b-1} & G_{b-1b+1}& .. & G_{b-1c} \\
  G_{b+1a+1} & G_{b+1b-1} & G_{b+1b+1}& .. & G_{b+1c} \\
  \vdots &  &  & & \vdots \\
  G_{c-1a+1} & G_{c-1b-1} & G_{c-1b+1}& .. &
  G_{c-1c}\end{array}\right), \quad
  {\bf X}=\left(\begin{array}{ccccccc}
  G_{ba} &  .. & G_{bb-1} & G_{bc+1} & .. & G_{bd-1}& G_{bc}\\
  G_{a+1a}& .. & G_{a+1b-1}& G_{a+1c+1} & .. & G_{a+1d-1}&
  G_{a+1c}\\
    \vdots &  &  & & &\vdots \\
    G_{b-1a}& .. & G_{b-1b-1}& G_{b-1c+1} & .. & G_{b-1d-1}&
  G_{b-1c}\\
  G_{c+1a}& .. & G_{c+1b-1}& G_{c+1c+1} & .. & G_{c+1d-1}&
  G_{c+1c}\\
   \vdots &  &  & & &\vdots \\
  G_{da}& .. & G_{db-1}& G_{dc+1} & ..& G_{dd-1}&
  G_{dc}
\end{array}\right)\notag
\end{eqnarray*}

\begin{eqnarray*}
{\bf S}=\left(\begin{array}{cccc}
  G_{aa} & G_{ab+1} & .. & G_{ac} \\
  G_{ba} & G_{bb+1} & .. & G_{bc} \\
  \vdots &  &  & \vdots \\
  G_{c-1a} & G_{c-1b+1} & .. & G_{c-1c}
\end{array}\right),\quad  {\bf Y}=\left(\begin{array}{cccccccc}
  G_{a+1a} & G_{a+1b} & G_{a+1a+1} & .. & G_{a+1b-1}& G_{a+1c+1} & ..& G_{a+1d-1}\\
 \vdots &  &  & & && &\vdots \\
  G_{b-1a} & G_{b-1b} & G_{b-1a+1} & .. & G_{b-1b-1}& G_{b-1c+1} & ..&G_{b-1d-1}\\
 G_{c+1a} & G_{c+1b} & G_{c+1a+1} & .. & G_{c+1b-1}& G_{c+1c+1} &..& G_{c+1d-1}\\
  \vdots &  &  & & & & &\vdots \\
G_{d-1a} & G_{d-1b} & G_{d-1a+1} & .. &
G_{d-1b-1}& G_{d-1c+1} & ..&G_{d-1d-1}\\
G_{ca} & G_{cb} & G_{ca+1} & .. &
G_{cb-1}& G_{cc+1} & ..&G_{cd-1}\\
G_{da} & G_{db} & G_{da+1} & .. & G_{db-1}& G_{dc+1} &..& G_{dd-1}
\end{array}\right)\notag
\end{eqnarray*}
In the above matrices, dots indicate continuous variation of the indices. In the absence
of dots, the indices are explicitly written and they may not change continuously. 
\end{widetext}

\end{document}